\documentclass{article}

\usepackage[latin1]{inputenc}
\usepackage{amssymb}
\usepackage{graphicx}
\usepackage{subfigure}

\usepackage{url}

\begin{document}

\title{Modeling browser-based distributed evolutionary computation systems} 

\author{ J. J. Merelo Guervós and Pablo García-Sánchez\\
  Dept. of Computer Architecture and Technology and CITIC\\
  University of Granada\\
  \texttt{(jmerelo,pablogarcia)@ugr.es}
}

\maketitle

\begin{abstract}
From the era of big science we are back to the ``do it yourself'', where
you do not have any money to buy clusters or subscribe to grids but
still have algorithms that crave many computing nodes and need them to
measure scalability. Fortunately, this coincides with the era of big
data, cloud computing, and browsers that include JavaScript virtual
machines. Those are the reasons why this paper will focus on two
different aspects of volunteer or freeriding computing: first, the
pragmatic: where to find those resources, which ones can be used, what
kind of support you have to give them; and then, the theoretical: how
evolutionary algorithms can be adapted to an environment in which nodes
come and go, have different computing capabilities and operate in
complete asynchrony of each other. We will examine the setup needed to
create a very simple distributed evolutionary algorithm using
JavaScript and then find a model of how users react to it by
collecting data from several experiments featuring different classical
benchmark functions. 
\end{abstract}

\section{Introduction}

The world has computational resources in spades. Most of them do not
belong to you or your lab. That does not mean you cannot use them. The
problem is how. 

Most theory in parallel computing has been devoted to predict and
optimize the performance in systems where the number of nodes, their
connections, and the time every one is dedicating to the computation is
known in advance. However, even if Big Science is not really over and
it is slated to come back, the era of
Citizen Science already started a few years ago (with SETI@home \cite{david-seti:home} and then BOINC \cite{boinc_grid04}) and it
offers a vast amount of computational resources to be used, if only
you know how to attract them. But there is a challenge: knowing, or at least having a
ballpark, of how your algorithm is going to perform in this uncertain
environment, where none of the factors is known: neither the number of
nodes, through how they are connected, to how long are they going to
be focused on doing what you want them to. That is why some effort is
needed to first understand the dynamics of the decision to participate
in an experiment that requires you to click on a link and then stay
for a while looking at the screen (or just leave it there running).

Besides, since Amazon started selling EC2 several years ago, reliable
and scalable computing resources are also available for a low price
and on demand. Recently, Google has also refurbished its offering
lowering their prices. This means that the conjunction of free or
low-cost cloud computing engines, volunteer computing systems and
untapped capability of desktop systems can be used for creating
massive, or at least potentially massive, distributed computing
experiments. These experiments can be easily created using open-source
repository sites like GitHub\footnote{\url{http://www.github.com}} and deployed automatically to Platform as
a Service (PaaS) products such as Heroku\footnote{\url{https://www.heroku.com/}} or OpenShift\footnote{\url{https://www.openshift.com/}}. 

In general, any volunteer computing experiment will have to be made
``in the open''. The fact that somebody is giving you, basically for
free, computing resources, means that the scientist using them has to
give back. The baseline is releasing the source code used: all
volunteer computing platforms, from SETI@home on, have done it. The
opposite is probably the reason why many companies like PopularPower \cite{buyya2001compute}
have folded or others like CrowdProcess have decided to turn their
product to in-house computation: there must be a mutual relationship
of trust among the scientist and the person that is running his/her
code in the browser. As it has been mentioned in the early stages of what
was then called {\em desktop grid computing} \cite{gc:bausch}, in
fact CPU cycle selling might not make economic sense since there are
not so much demand for it and supply is quite high. However, while
potential supply is in fact huge, {\em actual} supply will depend on
the person holding it willing to actually allocate it to a particular
company selling it or a particular experiment needing it, not to
mention the fact that the experiment {\em actually} has to draw the
attention of the supplier. That is why trust is essential, and using
free software might not be enough: Openness
has to progress from open source code to open science: releasing all
data obtained in the experiment in a repository such as GitHub,
mentioned above, and even allowing real-time access to experimental data
to users.


Another possible reason of the failure of former companies to create a
for-profit desktop grid might be the lack of a way to predict what is
going to be that supply in a particular moment. It is impossible to
predict, in advance, to know how many persons  are going to visit a
particular website. Even if you pull all the resources you have and
they lie across continents and time zones, the number of cycles
apportioned to a particular experiment will depend not only on the
users lending their web real state to the experiments (which is
usually the case for cycle brokerage companies) but especially to
users going to a web site and spending some time on them. Even if it
is theoretically impossible to predict to a high precision what
happens, it is in practice possible to approximate the number of
visits in a site, at least in a particular one, using time series. But
in the short term and using a more general model, there is still need
to model the behavior of users, so that more factors can be added to
the model other than the time series. This user behavior, of course,
presupposes that you are respecting their anonymity and privacy (not using
cookies, for instance) and that you are respecting the open approach
mentioned above. All computation can be done without the knowledge of
the user \cite{unwitting-ec}, but this would work against openness
that would, curiously enough, result in a huge decrease of the future
performance of any experiment you might want to perform. 

These are best practices that have been followed in the experiments in
this paper, that first presents the first versions of a platform for
distributed evolutionary algorithms (EAs) using the browser and a free (as
in free beer) backend, and second, shows the result from a
statistical point of view in order to put the basis for a model of the
metacomputer obtained by joining all volunteers and the free backend
used for the experiment. This is not an exhaustive or complete
exploration of the possibilities of this kind of computation, but it
is enough to present the free software used to perform the experiments
and will allow us to describe, in general, the behavior of the users
as well as the performance achieved on the experiments done so far,
which should show some advantage over doing the same kind of
experiments locally using available resources. 

The rest of the paper is organized as follows: next we will present
the state of the art in volunteer computing and its modelization. We 
will proceed to describe the experimental setup in Section
\ref{sec:exp}, some preliminary results in Section \ref{sec:res} and
finally we will present our conclusions in Section \ref{sec:conc}
along with future lines of work. 

\section{State of the art}
\label{sec:soa}

Using the web as a resource for distributed, or plainly user based,
evolutionary computation has a long history since JavaScript was
introduced as a browser-based language in 1997 and even before, when
other procedures, including Flash animations, VBScript, ActiveX or Java applets were
used. Java was pointed out in \cite{soares1998get} as a ``language for
internet'', providing some advantages such as multi-architecture compatibility or 
security mechanisms.
In that work, Soares et al. describe JET, a system that supports
the execution of parallel applications over the Internet. This system has 
a comprehensive statistics support, allowing its use for science, 
unlike the other systems compared in that paper.

However, Java (and all the other technologies)
is not universal in the
sense than an extra component, namely, the Java virtual machine, has
to be installed in the browser. JavaScript
\cite{flanagan2006javascript} was introduced in 1997 as a
browser-embedded language and has, since then, become a set of standards
\cite{ECMA-262} for the language, its components and future versions
that have been widely adopted by the industry and also by scientists,
who used them for creating a non-distributed EA on
the browser as early as 1998 \cite{jj-ppsn98}. 

The potential for volunteer computing using browsers was realized
later on \cite{sarmenta-bayanihan} as well as the potential for
mischief of users with code in their hands
\cite{sarmenta-sabotagetolerance}. However, these early efforts by
Sarmenta once again used Java and not JavaScript, making this effort
less universal. JavaScript can be used either for unwitting
\cite{unwitting-ec} or volunteer
\cite{langdon:2005:metas,gecco07:workshop:dcor} distributed
evolutionary computation and it has been used ever since by several
authors, including more recent efforts \cite{Desell:2008:AHG:1389095.1389273,duda2013distributed,DBLP:journals/corr/abs-0801-1210} that even
used the client's GPU \cite{duda2013gpu}. Many other researchers have
used Java \cite{chong:1999:jDGPi} and others have gone away from the
server-based paradigm to embrace peer to peer systems
\cite{jin2006constructing,10.1109/ICICSE.2008.99}. These computing
platforms avoid single points of failure (the server) but, since they
need a certain amount of infrastructure installed to start, the
threshold to join them is much lower. 

There have been relatively few efforts to analyze what is the
performance that can be obtained from these volunteer computing
effort. There was some attempt initially to dodge the issue by making
the algorithm adaptive to the kind of resources allotted to it
\cite{milani2004online}, which is actually not such a big problem in
algorithms such as the EA that can easily be
parallelized via population splitting or farming out evaluations to all
the nodes available. Lately, several approaches have focused on the
fault-tolerance of volunteer algorithms
\cite{gonzalez2010characterizing} which can, of course, be studied in
a more general distributed computing context
\cite{nogueras2015studying} or including it in a more general study of
performance of the EA itself 
\cite{DBLP:journals/gpem/LaredoBGVAGF14}. But the raw material of
volunteer computing, number of users and the time spent in the
computation in browser-based volunteer computing experiments, have only been studied in a limited way in
\cite{DBLP:journals/gpem/LaredoBGVAGF14} on the basis of a single
run. Studies using volunteer computing platforms such as SETI@home
\cite{javadi2009mining} found out that the Weibull, log-normal and Gamma distribution
modeled quite well the availability of resources in several clusters
of that framework; the shape of those distributions is a skewed bell
with more resources in the {\em low} areas than in the high areas:
there are many users that give a small amount of cycles, while there
are just a few that give many cycles. This is in concordance with the
results obtained in \cite{agajaj}.

As far as we know, this paper presents the only experiment that uses computational
resources that are as dissimilar as smartphones and powerful laptops
or desktop computers in a research center. The methodology used to
gather resources and the algorithms used will be described next. 


\section{Methodology}
\label{sec:exp}

In order to test the volunteer computing environment, a presentation
describing a low cost volunteer computing environment was
created. This presentation was actually delivered in several
conferences\footnote{Names withheld for the double blind review}. During the
conference, it was revealed to the persons attending it that they were participating
in the experiment. The same procedure was used when trying to gather
users in social networks: a description of the container (the
presentation) and disclosure of its purpose. The upper right corner of
the presentation shows the progress of the evolution. It
stops when the solution has been reached. 

Above the experiment, from the point of view of the user, has been
described. What it is actually running is an island model
\cite{muhlenbein1991parallel} in every browser that uses the server as
a {\em shuttle} to transfer individuals from one browser to another in
what can be eventually a fully connected topology; however, all
connections take place between the browser and the server.
This deals with
several problems. The connection is stateless: islands connect to the
server to send and receive a single individual but there is no task
balancing: all islands start clicking on an URL and finish when they
browse off to the next page. Fault tolerance is implicit, in the sense
that no island has information that cannot be, or generated, somewhere
else, although if the server fails the experiment results might be
lost. All operation is asynchronous, with islands entering and leaving
the experiment all of their own. 

We will next explain its different parts from the point of
view of the architecture itself and the algorithm that it is actually
running. First, we will describe the client code and next, in
subsection \ref{ss:server} the server architecture and where it is
hosted. 

\subsection{Volunteer distributed evolutionary computation in the
  browser}

As stated above, the evolutionary algorithms run mainly in the
browser. The problem run is a multimodal problem called {\em l-trap},
which has been used extensively as a benchmark for evolutionary
algorithms \cite{fernandes2009using,nijssen2003analysis}. This
function counts the number of bits in a sequence of $l$ and assigns
the local maximum if it has got 0 bits and the global maximum to $l$
bits. The fitness goes down to a {\em trap} as you increase the number
of bits, that is why it creates a rugged fitness landscape that is
difficult to surmount for evolutionary algorithms. Its difficulty is
increased as the number of traps concatenated grows, so that in some
cases it might need millions of evaluations to find the solution. In
our case we have used several values, from 30 to 50 traps, although we
were not so much focused on finding the solution (or the fastest way
of finding it) but on creating a experiment that could last for a
certain amount of time, around one hour, so that there would be a
chance for having many users checkout out the page and contributing
cycles to it. 

This fitness is implemented as part of the Nodeo evolutionary
algorithm library written in node.js, a server-based implementation of
JavaScript. However, since it is going to be run on the browser, a
procedure called ``browserification'' must be applied to it so that it
can run, without modifications, in any browser. The main difference
between JavaScript on the browser and node.js is the way they load the
external libraries. The algorithm used a population of 256
individuals, random initialization, a canonical evolutionary algorithm
with an elite of 2 individuals and crossover and then mutation applied
to all the population pool, that was generated using 2-individual
tournament selection. Once again, the particular evolutionary
algorithm used is not so important, since we were rather focused on
the user behavior, although it is quite clear that the selective
pressure of the algorithm will have an impact in its performance and
will play for or against its asynchrony in many different and complex
ways \cite{jj:2008:PPSN}. 

The initial code was only modified to represent the fitness in a chart
(that uses the {\tt chart.js} JavaScript library) so that the user has
a visual feedback of what is going on. The {\em migration} part is
also added: after every 100 generations, a single individual is sent
to the server and another request is made to the server of a single
and random individual. This makes the procedure less greedy, and also
decreases the probability of getting the same individual that was sent
some generations back. These two operations are made in {\em fire and
  forget} style, that is, without any error checking, so that if the server is
down, or indeed if the algorithm is run locally, it is not interrupted
and continues operation until the solution is found, a string with all
ones.  

As a summary, the only thing needed to run an evolutionary algorithm
in the browser is to add {\em migration} functions. To make the
operation of the algorithm independent of them is a plus, because it
makes the local algorithm more fault-tolerant. The algorithm running
on every browser has the same parameters and starts from a random
population when the page is loaded.

\subsection{Server side}
\label{ss:server} 

The server was also written using node.js and the {\tt express.js}
module and has three REST \cite{Castillo12REST} 
functions, that is, three functions that can be accessed from any
client, including the browser, via a request encoded in an URL and a
standard HTTP command. These functions are\begin{itemize}
\item {\tt GET random} returns a random, non-evaluated individual,
  from the pool. The pool is just an array containing chromosomes,
  that is initialized at the beginning of the experiment.
\item {\tt PUT one} puts a single individual in the pool. In REST
  conventions, {\tt PUT} is used to create a resource, that is why it
  is used, in this case, to add a new individual to the pool.
\item {\tt GET log} returns the experiment log so far. It is used by
  the researcher to gather data, and as a transparency measure so that
  anyone can use it. 
\end{itemize}

In fact, the program includes more logic to serve the static pages
that contain the talk in which it is included and the JavaScript/CSS
code also, but it all amounts to a small amount of code. This code can
be deployed, in principle, in any server running node.js, but for the
purposes of this experiment Heroku, a Platform as a Service with a
free tier, has been chosen. It could be also be deployed, with small
or no modification, to OpenShift or other similar PaaS. Far from being
a freeriding way, using free or very low cost servers can be a
sustainable way of performing massive distributed evolutionary
computation experiments \cite{Merelo2014,jj:idc:lowcost}.

Having the pages and the server in the same domain allows to work in
default Ajax mode, that does not allow cross-site requests. However,
these can also be disabled if needed. In this case, every domain holds
a single experiment, there is no multi-tenancy of experiments, so its
management is quite simple.

When the experiment is finished, that is, the solution has been found,
the server is restarted using  command line tools provided by
Heroku. Every experiment was first done during the talk and then
announced in Twitter several times until enough data was gathered.

As indicated in the introduction, every part of the experiment and
data gathered have been released as free software in the GitHub
repository \url{https://github.com/JJ/idc-keynote}. These results
will be analyzed in the next section.

\section{Results}
\label{sec:res}

The litmus test that all experiments should pass, that is, that they
work and found the solution, passed. 
In fact, the first version of the
algorithm (until January 2015) used too much memory on the browser,
which crashed after a while running. A total of 7 experiments were 
done with this version. After January, a new version of the
evolutionary algorithm with non-cached fitness was used which did not
crash the browser and was thus allowed to run as much time as
needed. That was the version used in a lightning talk at the beginning
of February. A total of 10 experiments were made with this second
version. In the first version, finding the solution took a few
minutes. The second needed around 20 minutes. In the logs used, it was
not checked that the algorithm had found the solution, but in most
cases it had. More systematic experimentation would have to wait until
the next version, since in this first one we were looking for some
kind of systematicity in the behavior of users and the experimental
apparatus is the minimum possible to have it running.
\begin{figure*}[htb]
        \centering
        \subfigure[First experiment]{
                \includegraphics[width=0.45\linewidth]{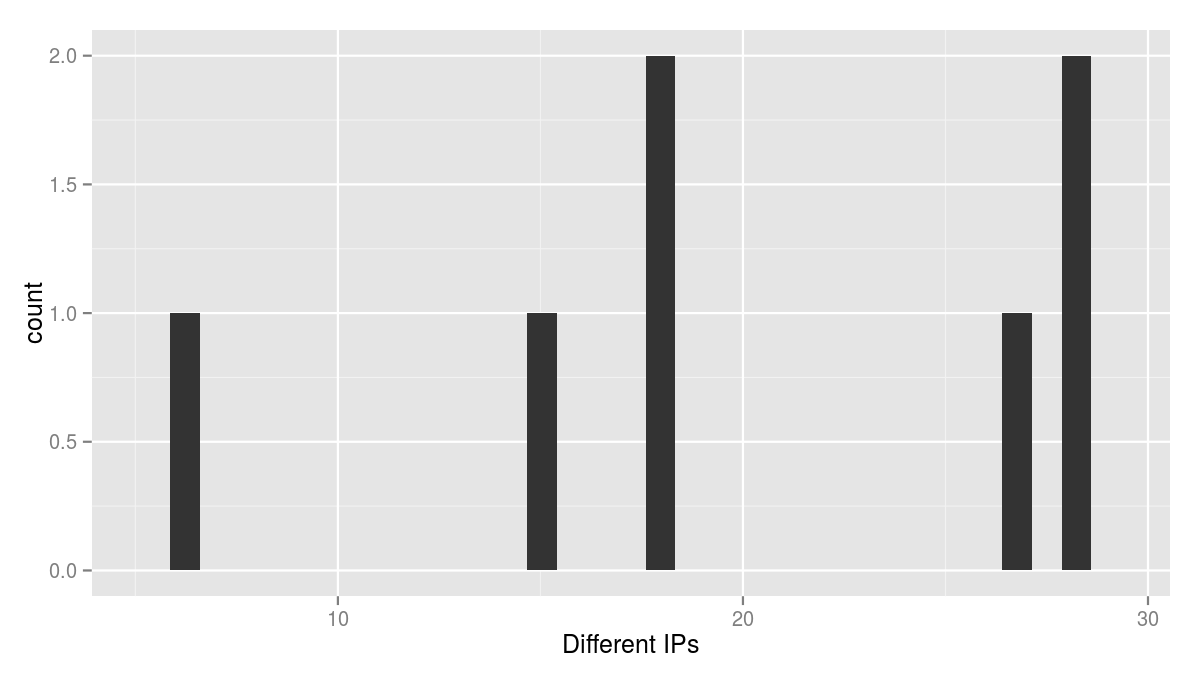}
                \label{fig:e1_p}
        }
        \subfigure[Second experiment]{
                \includegraphics[width=0.45\linewidth]{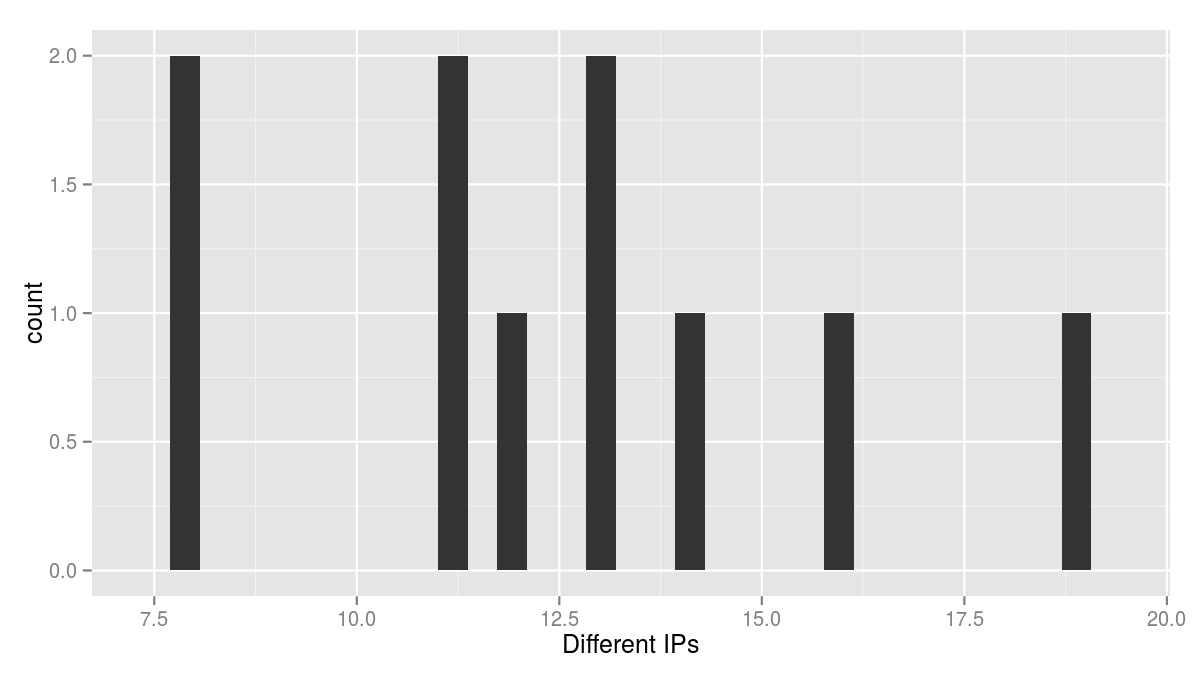}
                \label{fig:e2_p}
        }
        \caption{Histogram of number of different IPs in an experiment}\label{fig:histo:ips}
\end{figure*}

The fact that it run correctly during a massive lightning talk means that this setup is a
valid, no-cost, platform for massive evolutionary algorithms. There
were several ways of introducing it: during a talk, telling the
audience to visit the web that was posted in the first slide, and
after the talk, inserting the URL in a tweet with a brief explanation
and several tweets more indicating what it was about and what it
did. If any question arose, it was answered as fast as possible. In
the first experiment, the median number of different IPs was 18, with
a maximum of 28 and a minimum of 6. In the second, the median was 12.5
with a maximum of 19 and a minimum of 8. The histogram of different IPs
in each experiment is shown in figure \ref{fig:histo:ips}. Probably
the most remarkable thing is that a cluster of 6 computers can, as a
minimum, be gathered for a distributed computing experiment by just
announcing it as a tweet, but the fact that we can obtain more than a
dozen computers in more than half the cases is also remarkable, and
goes to prove the excess of computing resources that people,
willingly, lend to a simple experiment.
%
\begin{figure*}[htb]
        \centering
        \subfigure[First experiment, run 3]{
                \includegraphics[width=0.45\linewidth]{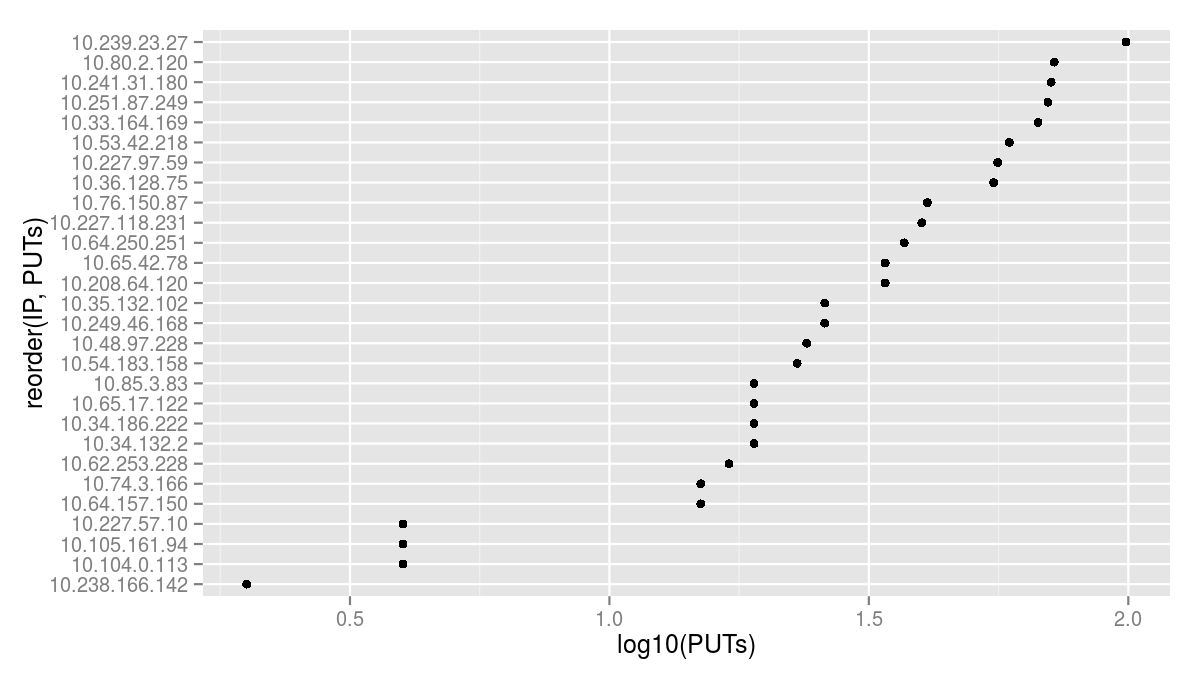}
                \label{fig:d1}
        }
        \subfigure[Second experiment, run 7]{
                \includegraphics[width=0.45\linewidth]{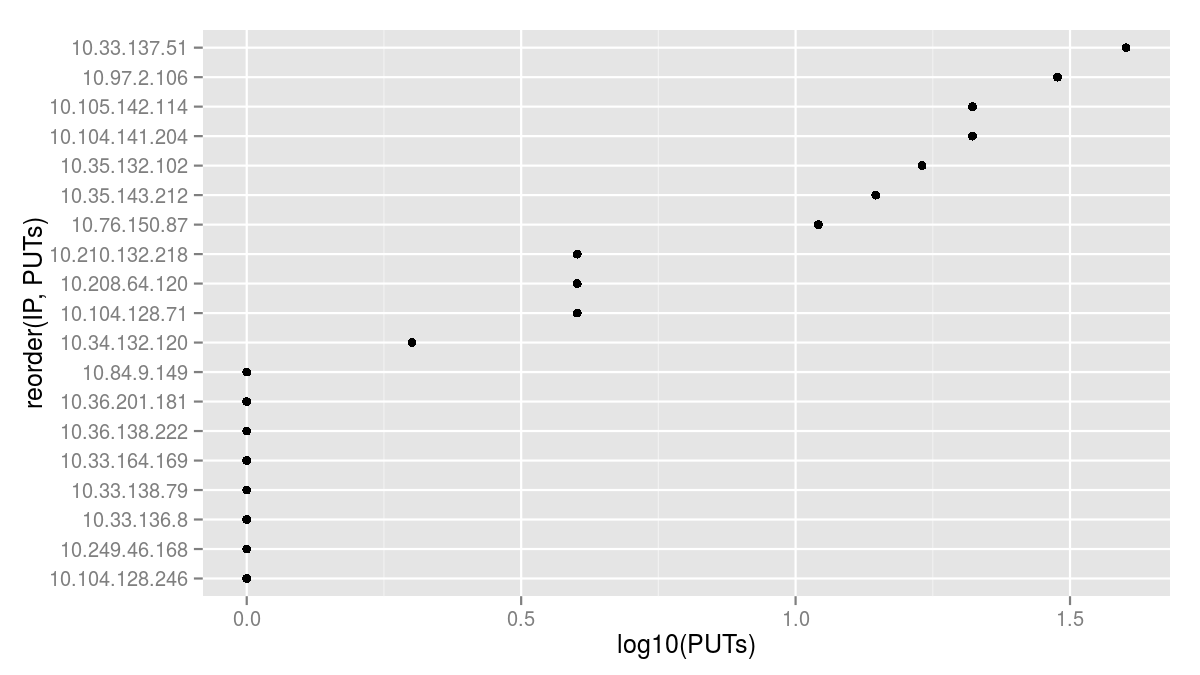}
                \label{fig:d2}
        }
        \caption{IPs and the number of {\tt PUT}s (one every $n$
          generations, 50 in the first experiment, 100 in the second
          experiment) they did, sorted. These correspond to the runs
          in each batch of experiments with the max number of
          different IPs. Please note the algorithmic scale in the $x$ axis.}\label{fig:dist}
\end{figure*}

Not all users contribute in the same way. The distribution for the
runs in each experiment with the most IPs is shown in Figure
\ref{fig:dist}, which chooses the experiment with the most number of
uses and ranks, tops to down, every IP with the number of PUTs (which
should be multiplied by the number of generations per PUT, 50 in the
first case, 100 in the second, although at this point it is not
important, since all nodes in the experiment do it after the same
number of generations) contributed to the experiment. We can observe
in both cases a power law, something already observed by \cite{agajaj}
and which implies that there is a fixed proportion between the number
of cycles (generations) contributed by the first and that contributed
by the second which is roughly the same as the one between the second
and third. Note also that, since the experiment is hosted in a PaaS,
the IPs listed are anonymous, with a 10 always as the first number.
\begin{figure*}[htb]
        \centering
        \includegraphics[width=0.8\linewidth]{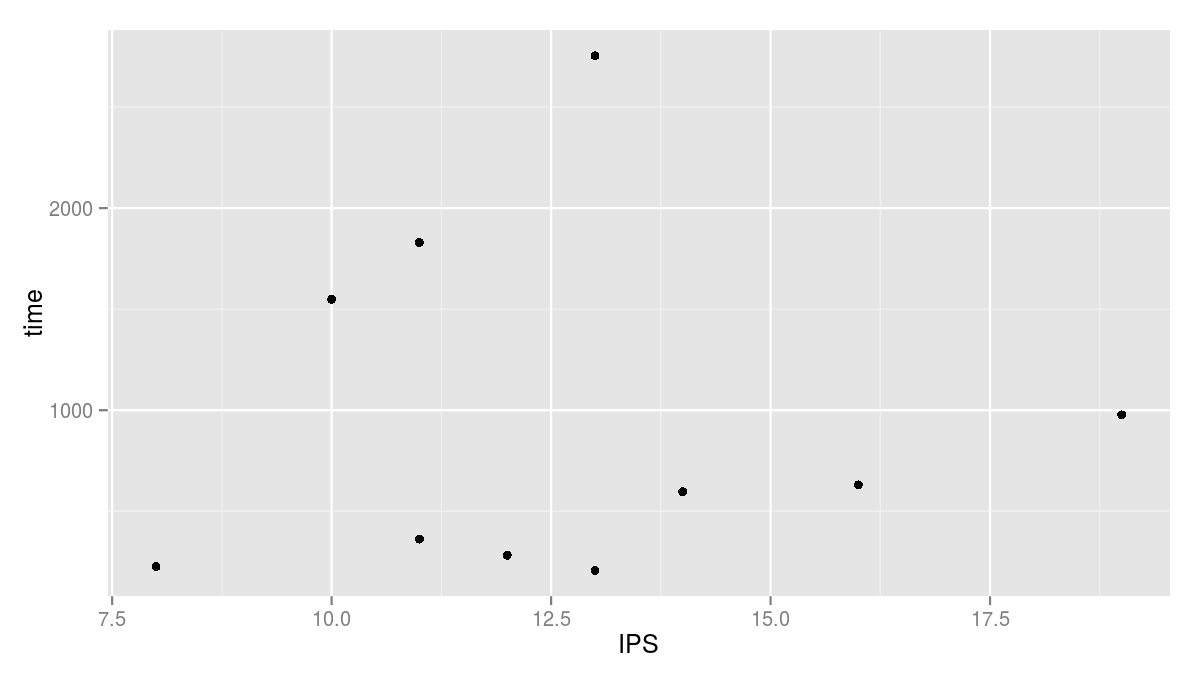}
        \caption{Total experiment time (time of the last PUT - time of
        the first one) vs number of IPs participating in it.}\label{fig:t}
\end{figure*}
\begin{figure*}[htb]
        \centering
        \includegraphics[width=0.8\linewidth]{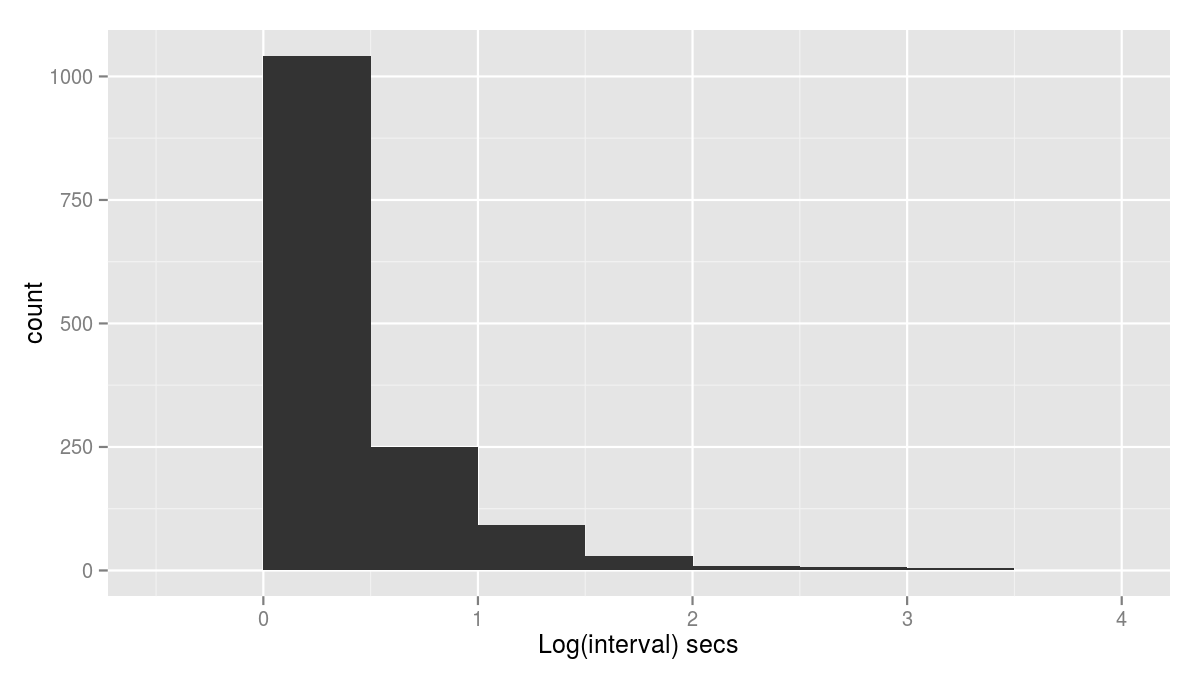}
        \caption{Histogram of the time between two PUTs for all the
          experiments in the second batch. The $x$ axis is logarithmic
        and bin size is 0.5, which means that the first bin contains
        all intervals of less than 3 seconds.}\label{fig:intervals}
\end{figure*}

At any rate, this exponential trend in the first posts breaks down
after a number of IPs. That could be due to several possible factors:
persons checking the web for just an instant, late comers, or very
slow computers that are not able to lend more cycles to the
simulation. The first IP in both cases contributes on the order of a
hundred PUTs, that is, several thousands generations and several
hundred thousands evaluations, which, once again, is quite remarkable
for such a simple experiment announced by an ephemeral tweet. It
should be noted that, since the speed in the browser virtual machines
(Chrome, mainly, but also Firefox) is as fast as the node.js version,
a parallel version does not add overhead to the single-browser
version, which, as mentioned above, can run in an independent
way: a second browser always adds to the first, even if the total
number of evaluations and the time needed will not speed up in the
same way. However, even taking into account the stochastic nature of a
single evolutionary algorithm, what we are going to find is a high
dispersion of the times needed to reach a solution, which are plotted
in Figure \ref{fig:t} against the number of IPs. There is a wide
range of durations, although there is no clear decrease of the time
needed as the number of IPs used increase. However, we will have to
investigate further the speedups achieved and if these are 
algorithmic or due to the particular asynchronous and web-based
implementation. Figure \ref{fig:intervals} might help understand the
kind of raw material we are using: It plots a histogram of the time
needed by all experiments to process 100 generations, 25600
evaluations and the associated evolutionary geneticry. The vast 
majority of browsers processes it in less than 3 seconds (logarithm( 
time ) < 0.5, which is equivalent to 3.16). However, it is interesting
to note that there are some intervals that go up to the thousands of
seconds, for which we cannot really offer an explanation, other than
the user leaving the page and returning after a while. However, 75\%
of the cases take less than 4 seconds, which is thus a statistical
measure of the kind of performance we can expect from clients for this
type of problem. We can also conclude that 75\%, or a majority, will
be in the fastest tier, although there will be a small amount of them
that will be quite slow and will probably be impossible to accommodate
in the distributed computing experiment.

\section{Conclusion}
\label{sec:conc}

In this paper we have presented our experience on using browser-based
computing applied to the design of evolutionary algorithms. In the
spirit of Open Science, we have released all material related to the
experiment, including this paper. Experiments have been performed at
several times of the day and announced in Twitter or during a live
presentation about distributed computation using the browser. 

What we conclude from these computers is that the nature of the
performance, which is due to the number of persons deciding to
participate in the experiment, is not totally random. There are at
least lower bounds we can count on: several computers (as few as six)
can almost always be relied on, and in some cases up to 30 can
participate in a single experiment. However, it is not clear how all
these computers contribute to the overall performance in terms of
time, although all experiments were performed until a solution was
found. Most computers participating in the experiment are in the same
performance tier, although a small percentage of them (which will be
around 25\%) are quite slow.

The evolutionary algorithm programmed in this way can easily
accommodate all kinds of browsers. However, we are using an homogeneous
configuration for all of them, which might not be the best from an
algorithmic point of view; a random parameter setting like the one
used by \cite{LNCS86720702} could offer better results, since we do
not know in advance what is the performance of the nodes. More
systematic experimentation is also needed, specially using different
kind of problems, including more complex problems in which the fitness
function is {\em heavier} and takes longer to be processed. However,
the main intention of this paper, which was to prove that distributed
evolutionary computation could be done efficiently in a volunteer/free
environment, has been sufficiently proved. 

There are many issues involved in using these resources: from adapting
current algorithms so that they match this environment 
to check which EA configuration works the best in it, or creating a
framework that can use it easily. But the main
challenge is that asking people to contribute resources implies that
you are opening your science to society and you have to give something
in return: you have to adopt a set of best practices that have come to
be known as Open Science, because ``Give, and it shall be given unto
you'', you will get as much back from society as you give to it
opening your science and explaining it to the public. This, among
other things, means that popularity will become directly performance
of the metacomputer you create by attracting more users. And this
interplay between the social network itself (popularity in Twitter,
people attending to the conference and interested in following the
slides in their computer) is another, very interesting, venue to
explore in the future. 

\section{Acknowledgments}

This work has been supported in part by TIN2011-28627-C04-02 and
TIN2014-56494-C4-3-P (Spanish Ministry of Economy and Competitivity),
SPIP2014-01437 (Direcci{\'o}n General de Tr{\'a}fico) and PYR-2014-17
GENIL project (CEI-BIOTIC Granada). 

\bibliographystyle{abbrv}
\bibliography{geneura-latin1,volunteer,javascript,ror-js,GA-general}

\end{document}